\documentclass[]{spie}  

\usepackage{amsmath,amsfonts,amssymb}
\usepackage{graphicx}
\usepackage[colorlinks=true, allcolors=blue]{hyperref}

\title{Simons Observatory Focal-Plane Module: Detector Re-biasing With Bias-step Measurements}

\author[a]{Yuhan Wang}

\author[b]{Tanay Bhandarkar}
\author[c,d]{Steve K. Choi}
\author[e]{Kevin T. Crowley}
\author[f]{Shannon M. Duff}
\author[a]{Daniel Dutcher}
\author[f]{John Groh}
\author[g]{Kathleen Harrington}
\author[a]{Erin Healy}
\author[h]{Bradley Johnson}
\author[i]{Jack Lashner}
\author[c,j]{Yaqiong Li}
\author[k]{Max Silva-Feaver}
\author[a]{Rita Sonka}
\author[a]{Suzanne T. Staggs}
\author[f,l]{Samantha Walker}
\author[a]{Kaiwen Zheng}

\affil[a]{Joseph Henry Laboratories of Physics, Jadwin Hall, Princeton University, Princeton, NJ 08544, USA}
\affil[b]{Department of Physics and Astronomy, University of Pennsylvania, 209 S 33rd St. Philadelphia, PA 19104, USA}
\affil[c]{Department of Physics, Cornell University, Ithaca, NY 14853, USA}
\affil[d]{Department of Astronomy, Cornell University, Ithaca, NY 14853, USA}
\affil[e]{Department of Physics, University of California, Berkeley, CA 94720, USA}
\affil[f]{National Institute of Standards and Technology, Boulder, CO 80305, USA}
\affil[g]{Department of Astronomy and Astrophysics, University of Chicago, Chicago, IL, USA}
\affil[h]{Department of Astronomy, University of Virginia, Charlottesville, VA 22904, USA}
\affil[i]{Department of Physics and Astronomy, University of Southern California, Los Angeles, CA 90089, USA}
\affil[j]{Kavli Institute at Cornell for Nanoscale Science, Cornell University, Ithaca, NY 14853, USA}
\affil[k]{Department of Physics, University of California San Diego, La Jolla, CA 92093 USA}
\affil[l]{Department of Astrophysical and Planetary Sciences, University of Colorado Boulder, Boulder, CO 80309, USA}

\begin{document} 
\maketitle

\begin{abstract}
The Simons Observatory is a ground-based cosmic microwave background survey experiment that consists of three 0.5 m small-aperture telescopes and one 6 m large-aperture telescope, sited at an elevation of 5200 m in the Atacama Desert in Chile. SO will deploy 60,000 transition-edge sensor (TES) bolometers in 49 separate focal-plane modules across a suite of four telescopes covering 30/40 GHz low frequency (LF), 90/150 GHz mid frequency (MF), and 220/280 GHz ultra-high frequency (UHF).  Each MF and UHF focal-plane module packages 1720 optical detectors spreading across 12 detector bias lines that provide voltage biasing to the detectors. During observation, detectors are subject to varying atmospheric emission and hence need to be re-biased accordingly. The re-biasing process includes measuring the detector properties such as the TES resistance and responsivity in a fast manner. Based on the result, detectors within one bias line then are biased with suitable voltage. Here we describe a technique for re-biasing detectors in the modules using the result from bias-step measurement. 
\end{abstract}


\section{Introduction}
The Simons Observatory (SO) is a suite of ground-based telescopes to be sited in the Atacama Desert in northern Chile at an altitude of 5200 m. SO will focus on measuring the temperature and polarization anisotropy of the cosmic microwave background (CMB) with over 60,000 transition-edge sensor (TES) bolometers spread among one 6 m large aperture telescope (LAT) and three 0.5 m small aperture telescopes (SATs)\cite{forcast}. Forty-nine separate universal focal-plane modules (UFMs) spanning six frequency bands from 30 GHz to 280 GHz host the TES bolometers and readout circuitry based on microwave SQUID multiplexers\cite{Dober_2021}. The 30$/$40 GHz low frequency (LF) UFMs use lenslet-coupled sinuous antennas, while the 90$/$150 GHz mid frequency (MF) and the 220$/$280 GHz ultra-high frequency (UHF) UFMs implement horn-coupled orthomode transducers\cite{{so_20},{Galitzki_2018}}. The SLAC Superconducting Microresonator RF (SMuRF) electronics serve as the room temperature readout electronics\cite{Henderson_2018}.

There are 1720 optical detectors and 36 dark bolometers spread among 12 bias lines for each MF and UHF UFM. A single voltage bias value is shared by $\sim 150$ TES bolometers that connect to each bias line. The setting of the voltage bias has an impact on the bolometer performance qualities such as stability, sensitivity and time constant. During observation, detectors are subject to varying atmospheric emission. If detectors are not re-biased accordingly, the change of optical loading will move the TESes into different region of their superconducting transition, hence influence their performance. In the worst case, the TESes will become normal or superconducting and lose their sensitivity. The detector re-biasing process should be fast such that normal observation can be resumed soon after. Detector bias level is normally adjusted every 1 $\sim$ 2 hours for other CMB telescopes located in the same area, such as ACT\cite{{Choi_2020},{Aiola_2020}} and POLARBEAR\cite{P_A_R_Ade_2014,polar_bear_2020}. We anticipate the frequency of detector re-biasing in SO will be at the same order of magnitude.

In Section $\ref{UFM}$ we discuss the thermal circuit of the UFM and the Joule heating generated in the TES bias circuitry. They will influence the re-biasing process. Section $\ref{considerations}$ discusses the considerations about where to bias detectors. We describe two common ways to characterize the TES superconducting transition: I-V curve and bias-step measurement, and we introduce a technique of using the result from bias-step measurement to re-bias detectors in Section $\ref{rebias}$.

\section{UFM thermal circuit and Joule heating}
\label{UFM}

\begin{figure}
\begin{center}
\includegraphics[width=1\linewidth, keepaspectratio]{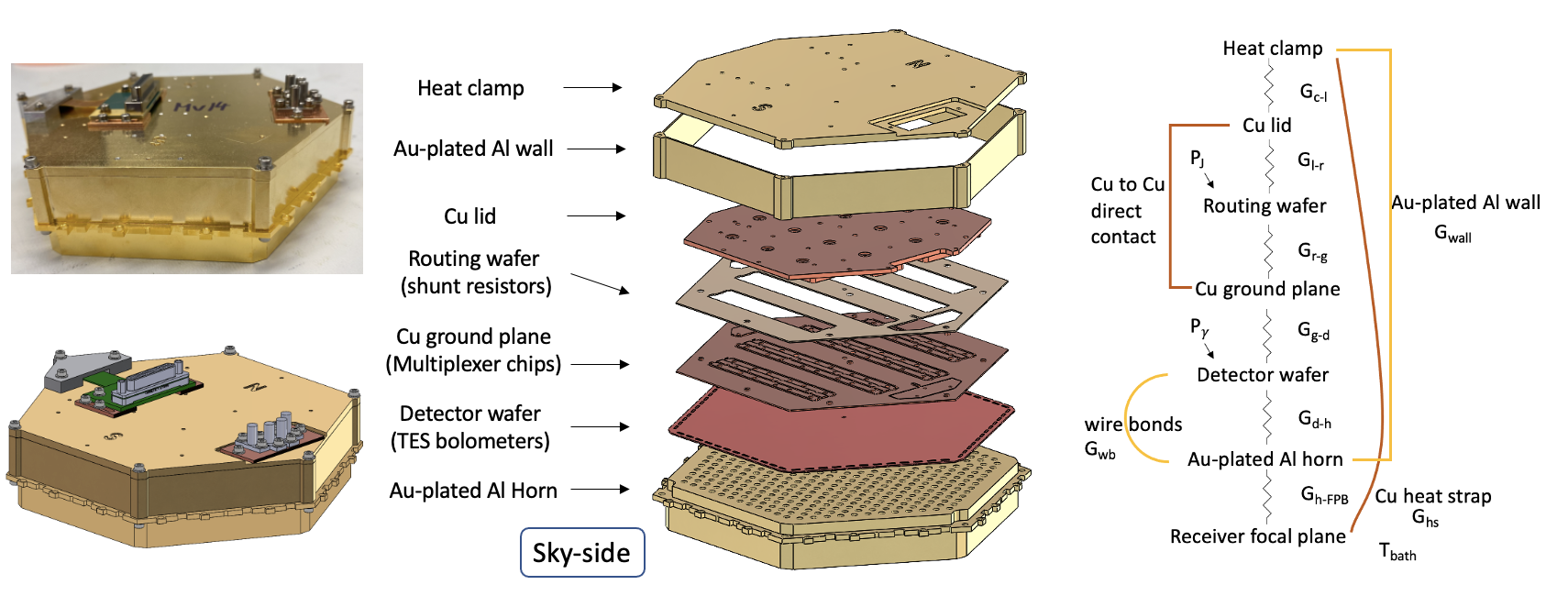}
\caption{Left: Picture and CAD model of the SO focal-plane module. Middle: Exploded view of the bulk parts in the SO focal-plane module. This picture is only showing some major bulk parts that are important for understanding the UFM thermal geometry. Right: UFM thermal circuit. During observation, the UFM will be bolted to the receiver focal plane. The receiver focal plane is considered as the UFM thermal bath. Copper heat straps connect the UFM Au-plated Al heat clamp to the focal plane and provide extra thermal path from the bath to the top of the UFM. The detector wafer is in the middle of a tightly stacked four wafer assembly, therefore we do not consider the thermal conductivity in between the layers individually. The copper lid is bolted to the copper ground plane and pushes the routing wafer down via pogo pins. The heat clamp provides clamping force through integrated tripods to all the layers and is connected to the horn via Au-plated Al wall. The wall is bolted to the clamp and the horn at each corner. The Joule heating $P_J$ from the shunt resistors is generated on the routing wafer and the photon power $P_\gamma$ is generated on the detector wafer. 
\label{fig:ufm.png}
}
\end{center}
\end{figure}

The UFM is a complicated assembly\cite{{McCarrick_2021},{Healy_2020}} with a complicated thermal circuit. In the UFM, the detector stack is placed atop the optical coupling. The detector stack consists four individual wafers, starting from the sky-side they are: the choke, the waveguide interface plate, the detector wafer and the backshort assembly. TES bolometers are located on the detector wafer. For MF and UHF UFMs, gold bonds are laid from the PdAu pads on the detector wafer to the Au-plated Al horn to provide an extra heat path. Above the detector stack, there is a copper ground plane hosting the multiplexer chips. The routing wafer locates above the copper ground plane and provides the 12 bias line and shunt resistors. A copper lid and Au-plated Al heat clamp stack above the routing wafer pushing wafers down to make sure a good thermal contact between each layer. The left panel of Figure $\ref{fig:ufm.png}$ shows a picture and a CAD model of the UFM.

During observations, UFMs will be mounted to the 100 mK focal plane with 7 UFMs in one SAT and three per LAT optics tube\cite{Galitzki_2018}. Copper heat strap will be added to connect the UFM heat clamp to the receiver 100 mK stage. The receiver focal plane is considered as the thermal bath for the UFM. The right panel of Figure $\ref{fig:ufm.png}$ shows a thermal circuit of the UFM when it is mounted on the receiver focal plane. The middle panel of Figure $\ref{fig:ufm.png}$ demonstrates the stacking structure of the UFM by showing the bulk parts that are important for understanding the UFM thermal geometry. 

\begin{figure}
\begin{center}
\includegraphics[width=1\linewidth, keepaspectratio]{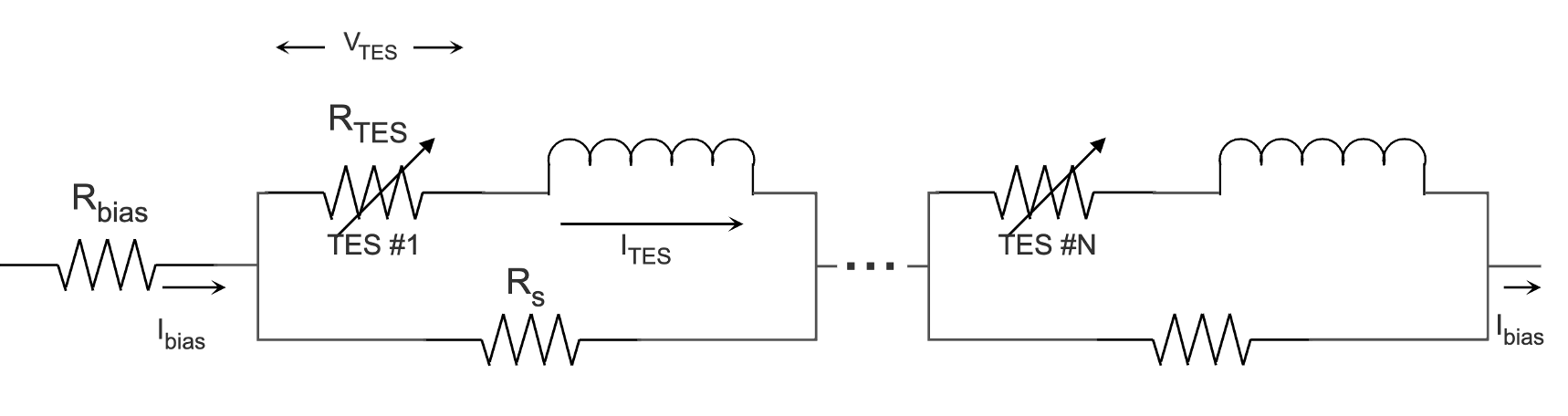}
\caption{TES bias circuitry diagram. Each MF and UHF UFM bias line hosts $\sim$ 150 TES bolometers. Each TES resistor $R_{TES}$ is in parallel with a shunt resistor $R_{s}$ and in series with an inductor that couples to a rf-SQUID. The SMuRF electronics provide the bias voltage $V_{bias}$ to each bias line and couple each bias line with warm stable bias resistors $R_{bias}$. The bias current $I_{bias}$, with amplitude $V_{bias}/R_{bias}$, flows into each TES bias circuitry and provides voltage bias $V_{TES}$ to each TES. The TES current $I_{TES}$ is read out by the coupled rf-SQUID.
\label{fig:tes_circuit.png}
}
\end{center}
\end{figure}

Each TES resistor (with normal resistance $R_n \sim 8~\mbox{m}\Omega$) is in parallel with a shunt resistor (with $R_s \sim 400~\mu\Omega$) so that it can be voltage-biased in its transition by bias current $I_{bias}$. A diagram from the TES bias circuitry can be found in Figure $\ref{fig:tes_circuit.png}$. 

The total Joule heating $P_J^{total}$ from the TES bias circuitry in the UFM can be expressed as:
\begin{equation}
P_J^{total} = P_J^{shunt} + P_J^{TES},
\end{equation}
in which the $P_J^{shunt}$ is from all the the shunt resistors and $P_J^{TES}$ is from the TES bolometers. When we bias the TESes, we first over-bias them normal for a short period of time before dropping onto the transition. When over-biasing TESes to their normal state, $P_J^{shunt}$ generated by the applied bias voltage is estimated to be over 360 nW and the $P_J^{TES}$ is over 18 nW for each UFM. Such heat will increase with increased drive normal bias voltage. As a comparison, during observation, $P_J^{total}$ is $\sim$ 77 nW when operating TESes around 50${\%}{R_n}$ and saturation powers for 90 GHz and 150 GHz detectors are less then 10 pW\cite{McCarrick_2021,Wang_2022}. One thermal path from the routing wafer to the bath goes through the detector wafer, hence $P_J^{shunt}$ will dissipate through the detector wafer. The thermal conductance between the routing wafer and the detector wafer is hard to tune. To control the influence of $P_J^{shunt}$ to the detector wafer, we can boost up other thermal conductance, for example, we can increase the number of copper heat straps connecting the heat clamp and the receiver focal plane. We can also reduce the amount of $P_J^{shunt}$ during UFM operation; for example, we can reduce the number of times driving detectors normal. More details about the UFM thermal performance and the dependence between the array heating and the UFM geometry will be discussed in future work.

\section{Considerations about where to bias detectors}
\label{considerations}
During observation, detectors are subject to varying atmospheric emission and hence need to be re-biased accordingly. At certain optical loading and bath temperature, the detector properties such as time constant and responsivity vary with the TES tuned depth into the superconducting transition. To read out the changing resistance of each voltage-biased TES, we measure the time-dependent current via an rf-SQUID ammeter. The normal resistance $R_N$ of each TES is measured in the laboratory prior the deployment\cite{Wang_2022}. A conceptually simple measure of the tuned depth into the superconducting transition is the $R_{TES}$ as a fractional percent of $R_N$: ${\%}{R_n}$. This is commonly used as a proxy for other metrics such as signal-to-noise and dynamic range. 

There are three main considerations about where to bias detectors: the array sensitivity, the detector time constant and the dynamic range. Section \ref{noise} discusses the expected detector noise and its behaviour when detector is in transition. This is related to the consideration about the detector sensitivity. When choosing the bias voltage for each bias line, we start from the detector level by choosing the suitable bias voltage for each detector and then move to the array level deciding the bias voltage for each bias line using a metric, this process is described in Section \ref{metric}. 

\subsection{Detector noise in transition and sensitivity}
\label{noise}
When detector is in transition the expected total noise can be expressed as:
\begin{equation}
NEP = (NEP_{\gamma}^2+NEP_{G}^2+NEP_{ro}^2+NEP_j^2)^{1/2},
\end{equation}
where the $NEP_{\gamma}$ stands for the bolometer noise from photon loading, $NEP_{G}$ stands for the thermal carrier noise, $NEP_{ro}$ stands for the readout noise and $NEP_{j}$ stands for Johnson noise.

For $NEP_{\gamma}$, we consider it coming from two parts: the fluctuation in the number of photons received by the detector, which follows Poisson statistics, \cite{{Richards_94},{Zmuidzinas_03}} and a second term recovering the Dicke equation for the sensitivity of a radiometer \cite{{Richards_94}}:
\begin{equation}
NEP_{\gamma}^2 = 2\int h\nu P_{\nu} d\nu + 2\int \frac{P_{\nu}^2}{N_{\nu}} d\nu,
\end{equation}
where the $P_{\nu}$ is photon loading absorbed by the detector at optical frequency $\nu$ and $N_{\nu}$ describes effective number of modes received by the detector. 

Detector $NEP_{G}$ can be expressed as\cite{Mather_82}: 
\begin{equation}
NEP_G^2 = 4k_BFT_c^2G,
\end{equation}
where the $k_B$ is the Boltzmann constant, F is a numerical factor, $T_c$ is the TES critical temperature and G is the bolometer thermal conductance. 

Both $NEP_{G}$ and $NEP_{\gamma}$ are independent of what ${\%}{R_n}$ the TES is at in the transition region.

The $NEP_{ro}$ describes the noise from components in the readout chain such as SQUIDs and amplifiers. The $NEP_{j}$ is generated by the thermal fluctuations of the charge carries in the detector bias circuitry. Both readout noise and Johnson noise come naturally in the form of noise equivalent current (NEI) but can be translated into NEP using:
\begin{equation}
\label{NEP_NEI}
NEP = \frac{NEI}{|s_i|},
\end{equation}
where $s_i$ stands for the TES responsivity\cite{Irwin2005}. When a detector is in transition, both $NEP_{ro}$ and $NEP_{j}$ are small comparing to $NEP_{\gamma}$ and $NEP_{G}$. Even though $NEP_{ro}$ and $NEP_{j}$ change with $R_{TES}$, their contribution to the total NEP is not significant.

From in-lab measurements, we have observed small variation of individual detector's NEP at different ${\%}{R_n}$ around the middle of the transition but larger variation of detector's NEP across the array as shown in the left panel of Figure \ref{fig:NEP.png}.

For each individual detector, the sensitivity $NET$ in unit of $\mu K\sqrt{s}$ is a function of total efficiency $\eta$, NEP and conversion factor $r$:
\begin{equation}
\label{NET_NEP}
NET = r^{-1}NEP / \eta,
\end{equation}
where $r$ converts CMB temperature fluctuations in $\mu K \sqrt{s}$ at input to $aW / \sqrt{Hz}$ at the detector. When deployed into receivers, $\eta$ for each detector is decided by the receiver optics and the detector optical efficiency. 

We measure the optical efficiencies of one third of the detectors in each UFM in laboratory with an internal cold load\cite{Wang_2022}. Another CMB experiment ACT has seen a large variation of detector optical efficiencies in the LF array from in-lab testing\cite{Li_2018} and on-site characterization\cite{Li_2021}. Although we have not observed such large variation of detector optical efficiencies across the array yet, we do not assume the detector optical efficiency will be a constant within one UFM when calculating array sensitivity in the field. Measurements of UFM optical efficiency will be reported in future work.

\subsection{Bias voltage choosing metric}
\label{metric}

Since each MF and UHF UFM has 12 bias lines and detectors on each bias line are biased under the same bias voltage, we consider the sensitivity in the unit of bias line. For each bias line, the sensitivity can be expressed as:
\begin{equation}
\label{NET_bl}
NET_{bl} = \sqrt{\frac{1}{\sum_{n=1}^{N} \frac{1}{NET_n^2}}} ,
\end{equation}
assuming there are N detectors on that bias line. When choosing the operating bias voltage for each bias line, we can consider minimizing $NET_{bl}$ by constructing a 2-d matrix that includes each detector's NEP at different bias voltages. However, constructing such matrix and minimizing Equation \ref{NET_bl} requires taking measurements at various bias voltages and is time consuming. Here we introduce a method to approximate the suitable operating bias voltage for each bias line.

We anticipate that we will re-bias detectors every 1 $\sim$ 2 hours similar to other CMB telescopes in the area. When choosing the bias voltage for each re-biasing process we first consider the detector stability as a constraint on choosing the lower bound of suitable ${\%}{R_n}$. Stable detector operating region will be measured in the laboratory\cite{Wang_2022} before deployment for each UFM. We then consider the fluctuations in the emission of water vapor and the elevation angle of the telescope. They will influence the optical loading during the next scan. Our goal is to bias each individual detector such that it can stay in transition with increased and decreased optical loading within one scan. Fluctuations in the emission of water vapor can be predicted from precipitable water vapor (PWV) forecast\cite{{Cort_s_2020},{apex}}. Studies from past CMB experiment located in the same area have demonstrated the impact of PWV to the data quality\cite{D_nner_2012,Hasselfield_2013} and have shown measurement of PWV around SO site is consistent between different measuring methods\cite{Morris_2022}. We explore the possibility of utilizing the PWV forecast to guide the choice of operating ${\%}{R_n}$. For example, if the PWV forecast shows rapid increasing PWV during next scan, we then would want to purposely bias detectors with low ${\%}{R_n}$, such that detectors will not saturate later at higher PWV. In this example, the upper bound of operating ${\%}{R_n}$ range is determined by the highest PWV from the forecast and the lower bound is determined by detector stability. The transfer function between the PWV and the detector operating range, which also depends on the total efficiency $\eta$, can be measured in the field. On going work in ACT is studying the fluctuation power response at different frequency band as a function of PWV.

\begin{figure}
\begin{center}
\includegraphics[width=1\linewidth, keepaspectratio]{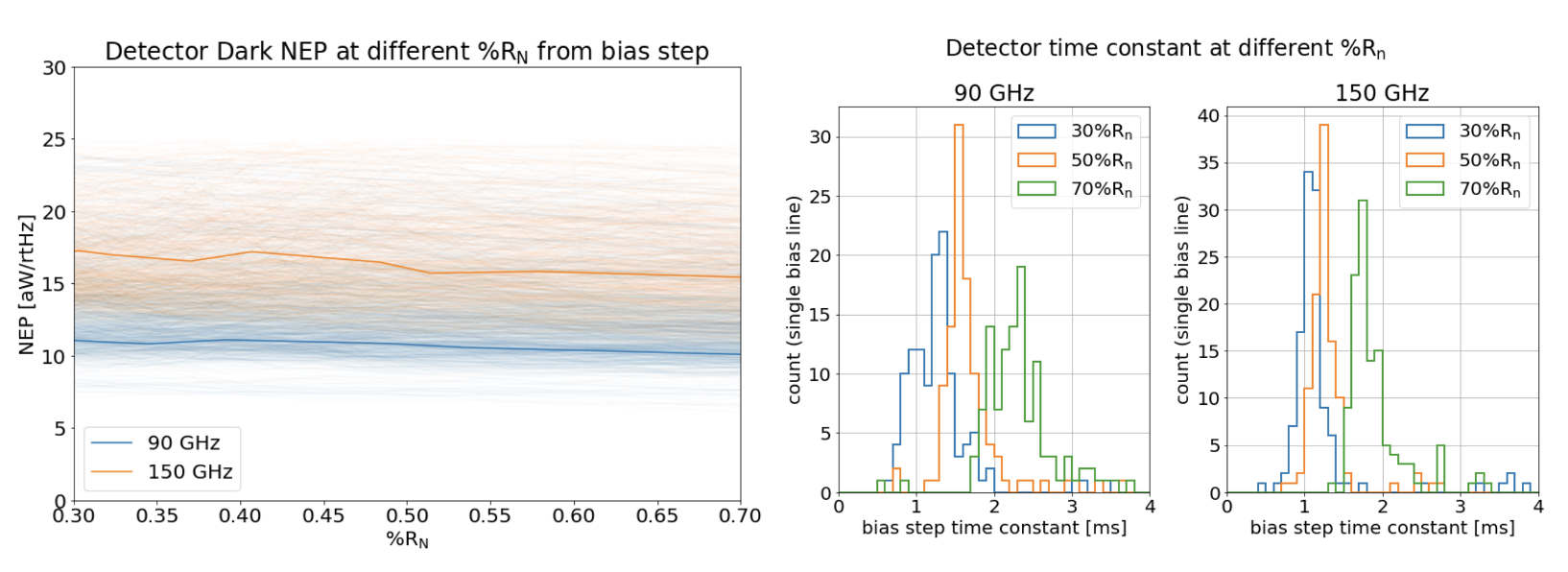}
\caption{Left: Detector NEP at different ${\%}{R_n}$ measured using the bias step method to extract the noise and responsivity. Showing the raw measurements using the flat region of a bias-step to extract NEI and calculating $s_i$ from the bias-step to transfer NEI into NEP. One 150 GHz and one 90 GHz detector are highlighted. Right: Detector time constant at different ${\%}{R_n}$. The measurements were made using bias-step at 100mK without optical loading on the detectors.  
\label{fig:NEP.png}
}
\end{center}
\end{figure}

After a range of suitable operating TES ${\%}{R_n}$ is determined for each scan, we then consider the sensitivity and time constant for each detector. As demonstrated in section \ref{noise}, the detector sensitivity is not sensitive to the bias point around the middle of the transition. The detector time constant acts as a low pass filter on the TES current response and increases at higher ${\%}{R_n}$. The right panel of Figure \ref{fig:NEP.png} shows the time constant measurements at different ${\%}{R_n}$. Therefore, within a suitable operating range, we choose to operate the detector at the lowest suitable ${\%}{R_n}$ to achieve smallest time constant. The suitable bias voltage for each detector is then determined to be the bias voltage that biases the detector to the lowest suitable ${\%}{R_n}$.

Next we define a metric for a bias line with total N detectors as:
\begin{equation}
\label{rebias_metric}
\sum_{n=1}^{N}\frac{|V_{bias}-V_n|}{NET_n^2},
\end{equation}
where $V_n$ is the voltage which will bring the nth detector to the target ${\%}{R_n}$ and $V_{bias}$ is the bias voltage applied to this bias line. Different detectors will have different $V_n$ because there are variations in detector properties, such as $\eta$ and $R_n$, among the array. This metric assigns a weight to each detector according to their NET. In order to favor detectors with low NET and minimizing $NET_{bl}$, $V_{bias}$ is chosen such that the metric reaches local minimum around the median of $V_n$.

\section{Re-bias method}
\label{rebias}
There are two common ways to characterize the TES superconducting transition: I-V curve and bias-step. In this section we introduce these two methods and discuss their advantages and limitations. We describe the method of using bias-step measurements to re-bias detectors using the metric described in Section \ref{considerations}. We note that a similar re-biasing procedure was used in the SPIDER experiment\cite{Rahlin_2014}.
\label{sec:rebias}  

\subsection{I-V curve}
\label{IV}
The I-V curve measurement is commonly used to determine the bias voltage for detectors. The advantage of the I-V curve method is that it provides a full measurement of the TES transition region. During an I-V curve, detectors are first overbiased into the normal state, the bias voltage is then reduced to step down TESes through their superconducting transition. The change of the TES current is tracked by the rf-SQUID in the multiplexer chip. The high bias voltage needed to bias TES into the normal state will result in large amount of Joule heating from the TES bias circuitry (approximately 378 nW when driving one whole array normal) hence a thermal shock to the system. If using the TES normal branch measurement to calibrate the DC offset of the TES current, TES needs to be overbiased with a range of high bias voltage to ensure there are enough data in the normal state and the corresponding Joule heating will be increased.

To reduce the amount of heating, one can extend the cool down wait time during an I-V curve or take I-V curve on one single bias line at a time. Both methods will result in a longer run time.

\subsection{Bias-step}
\subsubsection{Bias-step measurement}
\begin{figure}
\begin{center}
\includegraphics[width=1\linewidth, keepaspectratio]{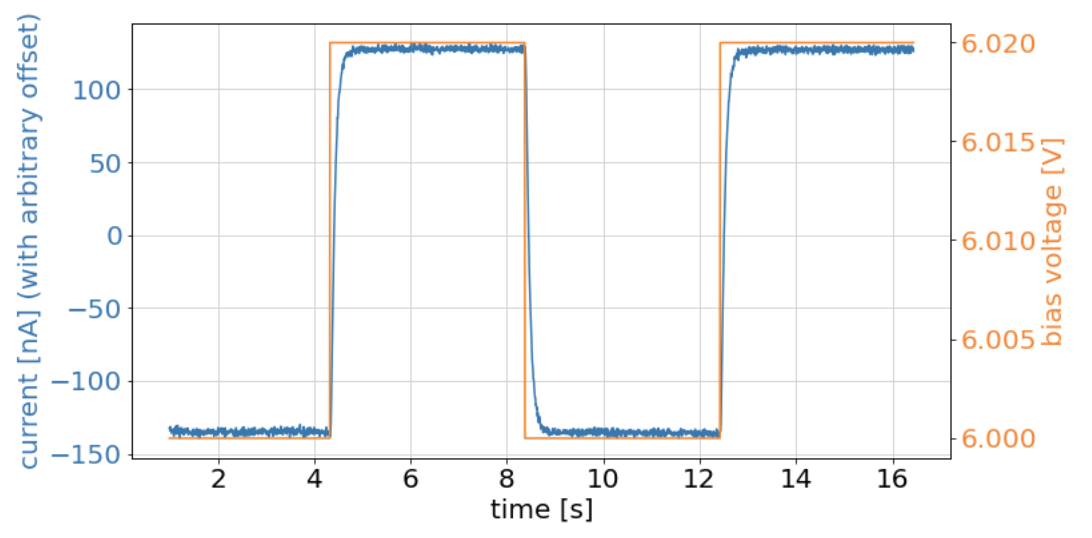}
\caption{An example of bias-step detector response. During this bias-step, a small-amplitude, in this case 0.02 V, square wave (orange) is played on top of a 6 V DC bias with a period of 8 s. The change of TES current (blue) is tracked by the multiplexer chip and is used to extract the TES resistance ($R_{TES}$) and the repsonsivity ($s_i$). The decay time of the detector response comes from a combination of the detector time constant and the warm bias circuitry low pass filter in SMuRF.
\label{fig:bias_step}
}
\end{center}
\end{figure}

A bias-step measurement is another way that can be used to select detector bias voltage. In a bias-step acquisition, a small-amplitude square wave is added to the DC bias level. The step function can be thought of as a two-point I-V curve. An example bias-step measurement can be found in Figure $\ref{fig:bias_step}$. The bias-step measurement does not require a measure of the TES normal state to calibrate the TES current DC offset. Instead, we assume that the TES bias power remains constant around the middle of the TES transition and derive TES current $I_{TES}$ using the change in bias current $\delta I_{bias}$ and the change in TES current response $\delta I_{TES}$\cite{{Niemack08},{grace2016}}. Such assumption can be verified in the laboratory using I-V curve measurements and has been found out to be valid when $R_{TES}$ is lower than 80${\%}{R_n}$. 

Figure \ref{fig:tes_circuit.png} shows a diagram of the TES bias circuitry. The bias current $I_{bias}$ can be calculated with commanded $V_{bias}$ and known $R_{bias}$ as: $V_{bias} / R_{bias}$. We then can express $I_{TES}$ as a function of bias power $P_{J}^{TES}$, $I_{bias}$ and $R_s$ as:
\begin{equation}
\label{I_tes}
I_{TES} = I_{bias} - I_{s} = I_{bias} - \frac{V_s I_{TES}}{R_s I_{TES}} = I_{bias} - \frac{P_{J}^{TES}}{R_s I_{TES}} = \frac{1}{2}(I_{bias} \pm \sqrt{I_{bias}^2 - 4\frac{P_{J}^{TES}}{R_s} } ).
\end{equation}
Taking the derivative of this and assuming $P_{bias}$ is a constant and independent of $I_{bias}$:
\begin{equation}
\frac{\delta I_{TES}}{\delta I_{bias}} = \frac{1}{2} (1 \pm \frac{I_{bias}}{\sqrt{I_{bias}^2 - 4\frac{P_{J}^{TES}}{R_s}}}).
\end{equation}
From the bias-step measurement we can measure $\delta I_{TES}$ and $\delta I_{bias}$, therefore we calculate $P_{J}^{TES}$ with measured or known quantities as:
\begin{equation}
P_{J}^{TES} = I_{bias}^2 R_s \frac{(\delta I_{TES}/\delta I_{bias})^2 - (\delta I_{TES}/\delta I_{bias})}{(1-2(\delta I_{TES}/\delta I_{bias}))^2}.
\end{equation}
We can then use $P_{J}^{TES}$ and Equation \ref{I_tes} to calculate $I_{TES}$. The $R_{TES}$ can then also be calculated and transfered into ${\%}{R_n}$ with measured $R_n$ from in-lab testing.

The step size of the bias step is small, hence the TES responsivity $s_i$ can be approximated in the small signal limit as \cite{Irwin2005}:
\begin{equation}
s_i = - \frac{1}{I_{TES}(R_{TES}-R_{s})} .
\end{equation}

The flat region of the bias-step measurement can be used to estimate the detector noise performance. Only the middle of the flat region is used to avoid the effect from the low pass filter in the warm SMuRF bias circuitry and detector time constant. The NEI can be approximated using the median of the amplitude spectral density of the time stream data between 5 and 50 Hz. This frequency range is selected to avoid both the low frequency region which can be contaminated by $1/f$ noise and the high-frequency roll-off defined by the anti-aliasing filter used in the SMuRF electronics. The NEP for each detector can then be calculated using Equation \ref{NEP_NEI}.

\subsubsection{Bias-step re-biasing}

We assume a situation that an initial I-V curve has been taken and detectors have been operating for some time at certain bias voltages, we now want to re-bias detectors for the next scan. Following the procedure described in Section \ref{metric}, a target operating ${\%}{R_n}$ for each TES is determined.

The bias-step re-biasing process measures resistances of TES bolometers in three different states. It starts with taking bias-step measurements on all 12 bias lines at the existing bias voltages. For each TES we call this the initial state with bias voltage $V_{bias,1}$ and measured TES resistance $R_{TES,1}$. Depending on whether the median of $R_{TES,1}$ of detectors is above or below the median of 50${\%}{R_n}$ in each bias line, a certain DC offset is applied to decrease or increase the current bias voltage on each bias line driving each TES into the intermediate state with $V_{bias,2}$. The second round of bias-step measurements are then taken, $R_{TES,2}$ for each TES is extracted. We then linear fit ($V_{bias,1}$, $R_{TES,1}$) and ($V_{bias,2}$, $R_{TES,2}$) to estimate the V-R relation around the middle of the transition for each TES. The approximated bias voltage needed to bias each detector to its suitable ${\%}{R_n}$ can then be predicted. 

From the bias-step measurements, the NEP of each detector can be calculated. We utilize the measured total efficiency from planet scan and in-lab optical efficiency measurement result to guide the conversion from the NEP into NET using Equation \ref{NET_NEP} and use the metric described in Equation \ref{rebias_metric} to find the most suitable bias voltage for each bias line. Each detector is then driven to the final state and a confirmation round of bias-step measurements is then taken at the decided bias voltage for each bias line.

\begin{figure}
\begin{center}
\includegraphics[width=1\linewidth, keepaspectratio]{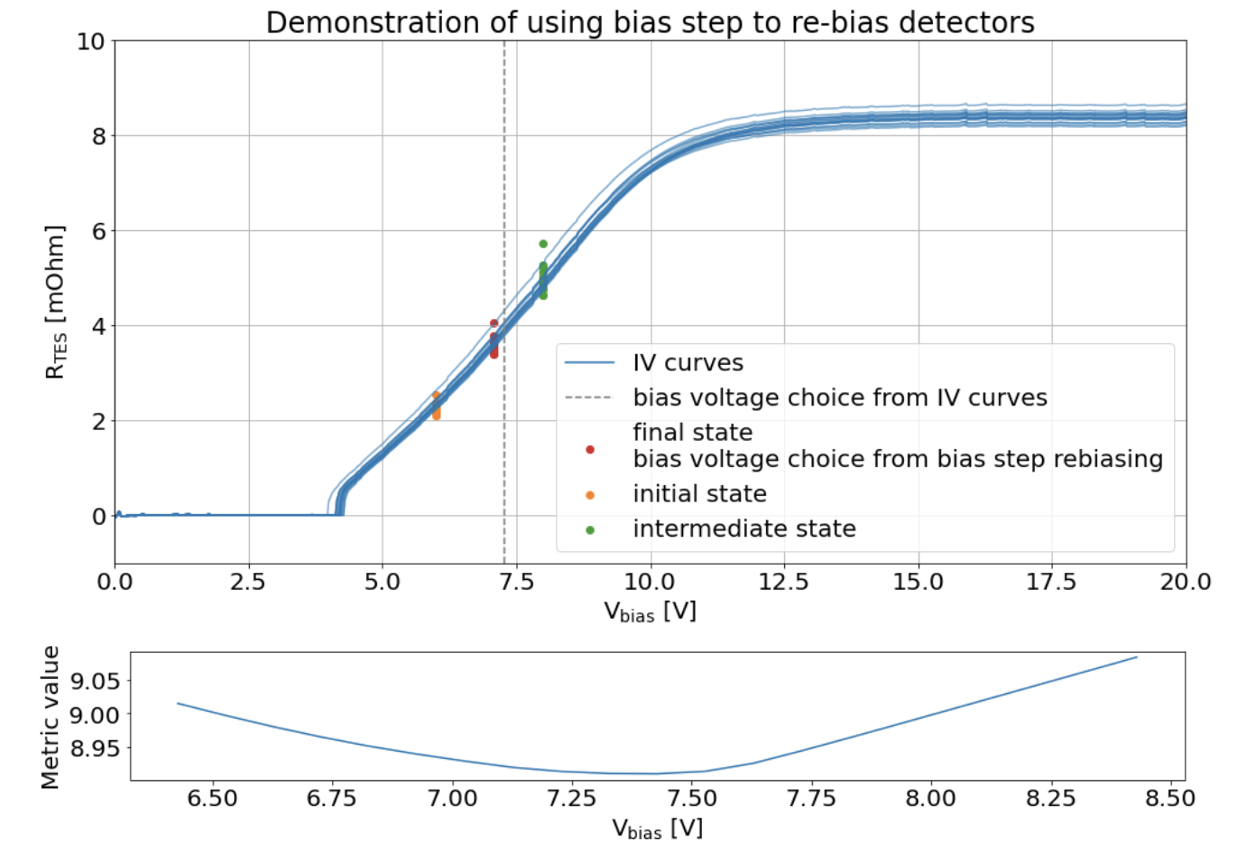}
\caption{In-lab demonstration of the bias-step re-biasing method and a comparison with the I-V curves. Showing data from detectors in one bias line without optical loading on them. Bias-step measurements were first taken at the initial state with the existing bias voltage. A DC offset was then applied to drive the detectors to the intermediate state. The target operating ${\%}{R_n}$ was set to be 50${\%}$ in this demonstration. From the two sets of measurements, the target $V_{bias}$ was calculated for each TES in this bias line. The algorithm then looked for local minimum of the metric described in Equation $\ref{rebias_metric}$. The optical efficiencies of the detectors were considered to be a constant for this in-lab demonstration as the measurements were made without optical loading. The final bias voltage was chosen and applied to drive the detectors to their final state. The top plot shows the three different states in this bias-step re-biasing process on top of a single bias line I-V curve taken on this bias line. The vertical dash line shows the bias voltage chosen from the I-V curves to target 50${\%}{R_n}$ without using the metric. The bottom plot shows the value of the metric described in Equation \ref{rebias_metric} at different bias voltages. We only consider the magnitude of the metric as the unit does not have physical meaning. In this demonstration the bias step re-biasing took $\sim$ 100 seconds to re-bias all 12 bias lines in the UFM.
\label{fig:demo.png}}
\end{center}
\end{figure}

Figure \ref{fig:demo.png} shows an in-lab demonstration of this method and a comparison with I-V curve measurements. Comparing to the I-V method, the bias-step method uses a smaller total bias voltage range and does not require the measurements of the normal branch. Therefore, bias-step provides a fast way to determine the re-bias voltages for all 12 bias lines simultaneously. In the demonstration, the total run time of the bias-step re-biasing was $\sim$ 100 seconds

\section{Conclusion and discussion}
In this paper, we discussed the considerations about where to bias detectors during observation. We introduced a method to determine bias voltage for each bias line, such that the total array NET can be minimized. We discussed the UFM thermal circuit and the Joule heating from the TES bias circuitry. 

We described two methods of characterizing the TES superconducting transition: the I-V curve and the bias-step. The I-V curve method provides a full measurements of the TES transition region. However it will introduce a thermal shock to the system when driving detectors normal. The bias-step provides an alternative way to measure and re-bias detectors. At small signal limit and under 80${\%}{R_n}$, bias step can be used to measure the $R_{TES}$, $s_i$, and detector NEP. We demonstrated using bias-step measurements and the metric we introduced to re-bias detectors. The in-lab demonstration of the bias-step re-biasing method took $\sim$ 100 seconds to re-bias all 12 bias lines in the UFM.

Further improvements of the detector re-biasing method for SO UFM will be explored. Detailed results of the UFM properties mentioned and used in this paper, such as optical efficiency and array sensitivity, and more discussions about the UFM thermal performance will be included in future work.

\begin{acknowledgements}
This work was supported in part by a grant from the Simons Foundation (Award 457687, B.K.) and private funding from universities.  
SKC acknowledges support from NSF award AST-2001866. 
YL is supported by KIC Postdoctoral Fellowship.
\end{acknowledgements}

\newpage

\bibliography{report} 
\bibliographystyle{spiebib} 

\end{document}